# Equilibria in the troposphere

F. Herrmann

*Abteilung für Didaktik der Physik, Universität Karlsruhe, D-76128 Karlsruhe, Germany*

We show that various types of equilibrium play an important part in the behaviour of the troposphere. In analogy to the electro-chemical potential (well-known in solid-state physics and electro-chemistry) a gravito-chemical potential and a gravito-thermo-chemical potential, as well as the corresponding equilibria are introduced. We shall show that
– the isothermal atmosphere is characterized by a constant gravito-chemical potential;
– the well-mixed or adiabatic atmosphere is characterized by a constant gravito-thermo-chemical potential.

Thus, a linear decrease of the temperature with the vertical coordinate corresponds to a state of equilibrium.

## I. ITRODUCTION

As a physicist who is mainly interested in the teaching of physics, I am working in a building which also accomodates a distinguished department of meteorology. It is near at hand, that I began to become interested in the physics of the atmosphere. Sometimes, however, I got into trouble when trying to understand my collegues from the other department. Obviouly, what they explained to me was consistent. Nevertheless, I often felt unable to translate what I had learnt from them into my own language. This is a natural phenomenon. Whenever two disciplines develop rather independently it happens that one and the same process or subject will be described with distinct models and different words.[1] With this article, I shall try to translate some of what I learnt from the meteorologists into a language which is more familiar to the physicist. In particular, I shall show that several phenomena in the atmosphere can be described as a process of establishing or of perturbing an equilibrium. The atmosphere is a peerless playing ground if one is interested in discussing all the facets of the concept of equilibrium. In my discussion, combined equilibria will play a particular role: equilibria of a type which is known by solid state physicists or physical chemists in the form of the electro-chemical equilibrium.

In section II we shall remember some basic concepts related to the establishment of equilibria. In section III examples of combinations of two potentials will be discussed. First we remember the electro-chemical potential and then we shall introduce a gravito-chemical potential. In section IV a "triple" potential will be considered: the gravito-thermo-chemical potential. This potential is the central concept of the article. Section V contains conclusions.

In this article we shall make the following simplifications:

1) We limit ourselves to the lower 10 km of the atmosphere, i. e., the *troposphere*.

2) We do not consider phase transitions, i. e., we only discuss the *dry atmosphere*.

3) We only consider atmospheres with a homogeneous composition, a condition which is realized within a good approximation in the troposphere of the earth.

4) Our most far-reaching restriction consists in the fact, that we assume that the values of all local quantities only vary in the vertical direction, or $z$-direction.

The symbols of physical quantities employed in this article have the following meaning:

$E$ = energy
$\sigma$ = entropy produced per time and per volume
$p$ = pressure

$V$ = volume
$T$ = temperature
$S$ = entropy
$R$ = gas constant
$c_p$ = molar specific heat at constant pressure
$\psi$ = gravitational potential
$m$ = mass
$g$ = acceleration of gravity
$\mu$ = chemical potential
$n$ = amount of substance
$\phi$ = electric potential
$Q$ = electric charge

Current densities are named $\boldsymbol{j}_X$, where $X$ is the flowing quantity. Thus, $\boldsymbol{j}_Q$ is the electric current density, $\boldsymbol{j}_S$ the entropy current density etc.

We need several "molar" quantities, i. e., extensive quantities divided by the amount of substance. We shall characterize them by a hat above the symbol of the extensive quantity. Thus, $\hat{S}$ is the molar entropy, $\hat{m}$ the molar mass etc.

We shall need several relationships from standard textbooks of thermodynamics:

1) The gravitational potential as a function of the height $z$ for distances from the earth's surface which are small compared to the radius of the earth:

$$\psi(z) = g \cdot z \tag{1}$$

2) The thermal equation of state of the ideal gas:

$$\hat{V} = \frac{RT}{p} \tag{2}$$

3) The molar entropy as a function of pressure and temperature for an ideal gas[2]:

$$\hat{S}(p,T) = R \ln \frac{p_0}{p} + c_p \ln \frac{T}{T_0} + \hat{S}(p_0,T_0) \tag{3}$$

4) The chemical potential as a function of pressure and temperature for an ideal gas[2]:

$$\mu(p,T) = RT \ln \frac{p}{p_0} - c_p T \ln \frac{T}{T_0}$$
$$+ \left( c_p - \hat{S}(p_0,T_0) \right)(T - T_0) + \mu(p_0,T_0) \tag{4}$$

In equations (3) and (4), $\hat{S}(p_0,T_0)$ is the molar entropy for an arbitrary reference pressure $p_0$ and an arbitrary reference temperature $T_0$, and $\mu(p_0, T_0)$ is the chemical potential for the same values of pressure and temperature.

The following numerical values will be needed:

$g = 9.81$ N kg$^{-1}$
$R = 8.3144$ J mol$^{-1}$ K$^{-1}$
$\hat{m} = 0.029$ kg mol$^{-1}$ (for dry air)
$c_p = 29.1$ J mol$^{-1}$ K$^{-1}$ (for dry air)
$\hat{S} = 198{,}7$ J mol$^{-1}$ K$^{-1}$ (for dry air at a temperature of 298 K and a pressure of 1 bar)

## II. DRIVING FORCES AND EQUILIBRIA

The Gibbs fundamental relation (or Gibbs fundamental form)

$$dE = \sum \xi_i dX_i$$

of thermodynamics[3,4] tells us that a system can change its energy in several ways. The number of terms is equal to the number of degrees of freedom of the system. In the couples of "energy-conjugated" variables at the right hand side of the equation, the first quantity, $\xi_i$, is, in general, an intensive quantity and the second one, $X_i$, an extensive quantity. Within the framework of thermodynamics one often considers systems whose Gibbs fundamental relation reads

$$dE = TdS - pdV + \sum \mu_i dn_i \tag{5}$$

The equation means, that the system under consideration, a gas for instance, can exchange energy in the following ways: in the form of heat $TdS$, by realizing work $-pdV$ or as chemical energy $\mu_i dn_i$. If the system is composed of only one substance and does not change its composition or phase, equation (5) simplifies to:

$$dE = TdS - pdV + \mu dn.$$

For other systems the Gibbs fundametal relation can contain many other terms of very distinct nature.[5] The system which has the following Gibbs fundamental relation

$$dE = TdS + \mu dn + \phi dQ + \psi dm$$

can exchange energy in the following forms: as heat, as chemical energy, as electric energy and as gravitational energy.

To each of the extensive quantities $X_i$, with the exception of the volume, corresponds a current with the current density $\boldsymbol{j}_{X_i}$. If the current is flowing through a "dissipator" or "resistor", i. e., if the current causes the production of entropy, the corresponding intensive quantity has different values at both ends of the dissipator. The energy dissipation rate per volume is[6]:

$$T\sigma = -\sum_i \mathbf{j}_{X_i} \cdot \operatorname{grad}\xi_i \qquad (6)$$

Here, $\sigma$ is the entropy production rate per volume. The expression on the right hand side of equation (6) is sometimes called the dissipation function. The gradient of $\xi_i$ can be interpreted as a driving force of the current $\mathbf{j}_{X_i}$. We first consider examples where only one single driving force is non-zero.

Imagine an isolated electric conductor in an initial state in which the electric potential varies from point to point. Equation (6) now reads

$$T\sigma = -\,\mathbf{j}_Q\,\cdot\,\operatorname{grad}\phi,$$

and the electric charge will redistribute until grad $\phi(z) = 0$ or

$\phi(z) = $ const.

This final state is a state of *electric equilibrium*.

Another example is a horizontal pipe which is provided with flow resistances and filled with air. The air is given an arbitrary initial pressure "profile" over the length of the pipe. According to equation (4) the chemical potential is also a function of the horizontal coordinate. Thus, there is a chemical potential gradient which causes a dissipative substance current, i. e., a redistribution of $n$ until

grad $(z) = 0$

or

$\mu(z) = $ const.

This final state is a state of *chemical equilibrium*.[7]

One might be inclined to interpret the concentration gradient as the driving force of the substance current, instead of the chemical potential gradient. In order to see that this is not appropriate consider a substance current in an inhomogeneous medium or "background". Imagine for instance, a bottle half-filled with water and the remaining volume containing carbon dioxide. At the beginning, the chemical potential of the $CO_2$ dissolved in the water is lower than that of the gaseous carbon dioxide. As a consequence, $CO_2$ will diffuse from the gaseous phase into the water. (The process can be accelerated by shaking.) At the end, $\mu(CO_2)$ has the same value everywhere, although the concentration in the solute phase remains lower than in the gaseous phase. Another example is well-known to every solid state physicist: The chemical potential of electrons depends on the conductor material.

*Thermal equilibrium* is defined in an analogous way: In the state of thermal equilibrium we have

grad $T(z) = 0$

or

$T(z) = $ const

and in the state of *gravitational equilibrium* there is

grad $\psi(z) = 0$

or

$\psi(z) = $ const.

There are other equilibria but these don't play a part in our considerations about the atmosphere.[8]

## III. TWO COMBINED DRIVING FORCES AND THE RESULTING EQUILIBRIA

### A. The electro-chemical equilibrium

Often one extensive quantity is coupled more or less tightly to another one. In general, this coupling is described by the phenomenological coefficients of Onsager's theory. Since we restrict our discussion to the special case of a completely tight coupling, we can do without this theory.

As a first example, let us consider the familiar case of electrons in a solid. For electrons, mass $m$ and amount of substance $n$ are tightly coupled. Moreover, there is a tight coupling between mass and electric charge $Q$, and therefore, also between amount of substance and electric charge. On the contrary, the coupling between entropy and the other extensive quantities is very loose. In the case of a tight coupling, the equilibrium takes upon a new character. Let us assume, that always $T(z) = $ const and $\psi(z) = $ const. These conditions are easy to meet. If the electrons are flowing in a conductor of constant temperature, the temperature of the electrons is also independent of the position. The gravitational potential does not play a part anyway because of the weakness of the gravitational field.

Electric charge and amount of substance are coupled via the molar charge (an integer multiple of the Faraday constant):

$$Q = \hat{Q}\cdot n$$

A similar relationship holds for the current densities $\mathbf{j}_Q$ and $\mathbf{j}_n$:

$$\mathbf{j}_Q = \hat{Q}\,\cdot\,\mathbf{j}_n \qquad (7)$$

Equation (6) now becomes:

$$T\sigma = -\mathbf{j}_Q \cdot \text{grad } \phi - \mathbf{j}_n \cdot \text{grad } \mu .$$

With (7) we get

$$T\sigma = -\mathbf{j}_n \cdot (\hat{Q} \cdot \text{grad } \phi + \text{grad } \mu)$$

The dissipation ceases, i. e., there is equilibrium, when

$$\hat{Q} \cdot \text{grad } \phi + \text{grad } \mu = 0$$

Because of the coupling between $Q$ and $n$, an equilibrium corresponding to only one of the exchange quantities $Q$ and $n$ cannot establish. Establishing the equilibrium for one of these quantities would in general cause a state of non-equilibrium for the other.

Since $\hat{Q}$ is independent of the position, the last equation simplifies to

$$\text{grad } (\hat{Q} \cdot \phi + \mu) = 0$$

and we can define a new, combined potential, the *electro-chemical potential*:[5,9]

$$\eta = \hat{Q} \cdot \phi + \mu$$

A gradient of the electro-chemical potential acts as the net driving force for the couple $[Q; n]$ of extensive quantities. In the case of *electro-chemical equilibrium* we have

$$\text{grad } \eta(z) = 0$$

or

$$\eta(z) = \text{const.}$$

An example for a situation in which the electro-chemical potential is constant, although the electric and the chemical potential taken separately are not constant, is the p-n-junction in equilibrium, familiar to every semiconductor physicist. A chemical potential gradient tries to pull the electrons from the n- to the p-region, whereas the electric potential gradient tries to pull in the opposite direction.

## B. The gravito-chemical equilibrium

Another important combined equilibrium is one which we propose to call the gravito-chemical equilibrium. Consider the air in the gravitational field of the earth. We first assume the atmosphere to be "isothermal", i. e., the temperature is independent of the height $z$:

$$T(z) = \text{const.}$$

In this case, two potential gradients are pulling at the air: the chemical potential gradient is pulling in the direction of decreasing pressure, i. e. upwards, the gravitational potential gradient is pulling downwards. Here too, both pertinent extensive quantities are tightly coupled:

$$m = \hat{m} \cdot n$$

The corresponding currents are related by the same factor:

$$\mathbf{j}_m = \hat{m} \cdot \mathbf{j}_n .$$

In this case equation (6) reads:

$$T\sigma = -\mathbf{j}_n \cdot (\hat{m} \cdot \text{grad } \psi + \text{grad } \mu)$$

and the condition for equilibrium is:

$$\hat{m} \cdot \text{grad } \psi + \text{grad } \mu = 0.$$

Since $\hat{m}$ is independent of the position this simplifies to

$$\text{grad } (\hat{m} \cdot \psi + \mu) = 0.$$

Again, there will be no equilibrium regarding one single extensive quantity and it is appropriate to define a new, combined potential, the *gravito-chemical potential*:

$$\gamma = \hat{m} \cdot \psi + \mu. \tag{8}$$

A gradient of the gravito-chemical potential acts as the net driving force for the couple $[m; n]$ of extensive quantities. In the case of *gravito-chemical equilibrium* we have:

$$\text{grad } \gamma(z) = 0$$

or

$$\gamma(z) = \text{const.} \tag{9}$$

With (1) and with (8), equation (9) transforms into

$$\mu(z) + g \cdot \hat{m} \cdot z = \text{const}$$

or

$$\mu(z) - \mu(0) = -g \cdot \hat{m} \cdot z. \tag{10}$$

Thus, the chemical potential decreases linearly with the height $z$. This is true for every fluid. Choosing the origin of the chemical potential such that $\mu(0) = 0$ and putting for $g$ and $\hat{m}$ the numerical values, we get

$$\mu(z) = -\frac{0.28\text{J}}{\text{m} \cdot \text{mol}} \cdot z .$$

If in equation (10) we insert the $\mu$-$p$-relationship of the ideal gas (equation (4) with $T = T_0$) we get the

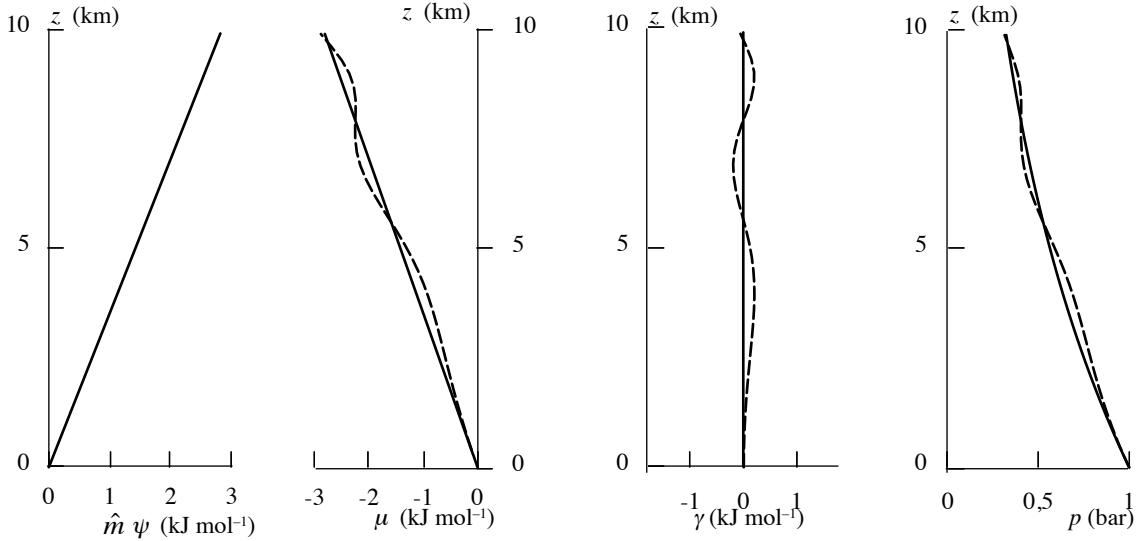

Fig. 1. The isothermal atmosphere. The dashed lines correspond to an initial state, in which the gravito-chemical equilibrium is perturbed. The full lines represent the equlibrium state. In this final state, the chemical and the gravitational driving forces balance each other.

barometric formula:

$$p(z) = p_0 \exp\left(-\frac{g\hat{m}z}{RT_0}\right).$$

Figure 1 resumes our results graphically. It shows the gravitational potential (in units of $1/\hat{m}$), the chemical potential, the gravito-chemical potential and the pressure as a function of the vertical coordinate $z$. The dashed lines correspond to an initial state in which the gravito-chemical equilibrium has not yet established. The full lines represent the final equilibrium state. The temperature is not shown because it is supposed to be constant. We had supposed that there is never a deviation from thermal equilibrium.

## IV. THREE DRIVING FORCES

### A. The hydrostatic equilibrium

In the preceding section we have admitted that the temperature over the height is constant, i. e., there is thermal equilibrium at the beginning, and the gas remains in thermal equilibrium even though the pressure is changing with time. In order to fulfill this condition, the gas should be a good thermal conductor. As a matter of fact, this assumption is completely unrealistic. The thermal conductivity of the air is so poor that just the opposite turns out much more reasonable: the thermal conductivity of the air is zero or, in other words, entropy and amount of substance (and mass) are tightly coupled. Under these circumstances the gas suffers the effect of three driving forces: the gradients of the gravitational potential, the temperature and the chemical potential.

Since both, the mass and the entropy, are tightly coupled to the amount of substance, we have

$$m = \hat{m} \cdot n$$

and

$$S = \hat{S} \cdot n$$

This implies for the current densities

$$\mathbf{j}_m = \hat{m} \cdot \mathbf{j}_n$$

and

$$\mathbf{j}_S = \hat{S} \cdot \mathbf{j}_n,$$

and equation (6) reads now

$$T\sigma = -\mathbf{j}_n \cdot (\hat{m} \cdot \operatorname{grad} \psi + \hat{S} \cdot \operatorname{grad} T + \operatorname{grad} \mu).$$

Again, there is no equilibrium regarding one single extensive quantity. The dissipation ceases, i. e. there is equilibrium, when

$$\hat{m} \cdot \operatorname{grad} \psi + \hat{S} \cdot \operatorname{grad} T + \operatorname{grad} \mu = 0. \tag{11}$$

Transforming equation (11) we shall meet an "old friend". As mentioned in the introduction we shall suppose that our field variables only change in the vertical, i. e. $z$-direction. Hence, the gradient can be substituted by the derivative $\partial/\partial z$. In equation (11) we replace $\psi$ with the aid of (1), $\hat{S}$ by means of equation (3) and $\mu$ with (4) and we get

$$g\hat{m} + \frac{\partial p}{\partial z}\hat{V} = 0$$

With the thermal equation of state of the ideal gas (2) we finally can write

$$\rho \cdot g + \frac{\partial p}{\partial z} = 0 , \tag{12}$$

the familiar equation for hydrostatic equilibrium. The surprise is not the result, but the way in which it was obtained. Indeed, we see, that the equilibrium state of the atmosphere which we just have discussed can not only be considered as a state of a mechanical equilibrium of forces, but also as a state in which the chemical, the thermal and the gravitational driving forces add to zero. For the moment this statement seems to be no more than a gimmick. In the following section, however, it will get some more importance.

We conclude that if an arbitrary pressure profile and an arbitrary temperature profile are applied to a gas and the gas is left to itself, a well-defined pressure profile and a well-defined temperature profile will establish. These profiles are the result of the common action of the gravito-chemical and the thermal potential gradients. In general, in such an equilibrium state non of these potentials is constant, i. e. independent of $z$.

## B. The gravito-thermo-chemical equilibrium

Let us now have a closer look at equation (11). The relation suggests to define a "triple" potential:

$$\varepsilon = \psi \cdot \hat{m} + T \cdot \hat{S} + \mu \tag{13}$$

Unfortunately, $\varepsilon$ cannot yet play the role of a driving force, as we shall recognize immediately. We made the assumption that the coupling between entropy and amount of substance is tight. However, we did not exclude, that the value of $\hat{S}$ is a function of the position $z$. Therefore, for the gradient of $\varepsilon$ we get

$$\text{grad}\,\varepsilon = \hat{m} \cdot \text{grad}\,\psi + \hat{S}\,\text{grad}\,T + T\,\text{grad}\,\hat{S} + \text{grad}\,\mu ,$$

i. e. a relation which not only contains the gradients of the three intensive variables $\psi$, $T$ and $\mu$, but also the gradient of the molar entropy .

We see, however, that the equation

$$\text{grad}\,\varepsilon(z) = 0$$

is identical with the equilibrium condition (11) if the molar entropy $\hat{S}$ is independent of $z$, i. e. if

$$\hat{S}(z) = \text{const} \cdot \tag{14}$$

An atmosphere for which this relation holds is called an adiabatic atmosphere. Actually, there is a simple means to fulfill this condition. The atmosphere has to be intermixed, and that happens always when turbulent currents are present. Indeed, for the well-mixed atmosphere the gradient of the potential $\varepsilon(z)$ defined by equation (13) is the pertinent driving force. We call $\varepsilon(z)$ the *gravito-thermo-chemical potential* or GTC potential. The equilibrium for which

$$\varepsilon(z) = \text{const}$$

or

$$\text{grad}\,\varepsilon(z) = 0$$

is the *gravito-thermo-chemical equilibrium*.

If (14) is fulfilled, equation (3) provides a relation between pressure and temperature:

$$R \ln \frac{p}{p_0} = c_p \ln \frac{T}{T_0} \tag{15}$$

which is one of the familiar adiabatic equations.[10]

Combining (4) and (15) we get the chemical potential as a function of the temperature:

$$\mu(T) - \mu(p_0, T_0) = \left[ c_p - \hat{S}(p_0, T_0) \right](T - T_0) . \tag{16}$$

Inserting (16) into equation (13), and taking into account condition (14) we get the GTC potential as a function of the vertical coordinate and the temperature:

$$\begin{aligned}\varepsilon(z,T) &= \hat{m}\psi(z) + \hat{S}T + \mu(T) \\ &= g\hat{m}z + c_p(T - T_0) + T_0 \cdot \hat{S}(T_0, p_0) + \mu(T_0, p_0).\end{aligned} \tag{17}$$

We now assume GTC equilibrium, i. e.,

$$\varepsilon(z) = \text{const}.$$

With equation (17) we find

$$T(z) = T(0) - \frac{g\hat{m}}{c_p}z \tag{18}$$

or, using the numerical values of $g$, $\hat{m}$ and $c_p$:

$$T(z) = T(0) - 0.0098\,\frac{\text{K}}{\text{m}} \cdot z ,$$

i. e. a linear decrease of the temperature with the height.

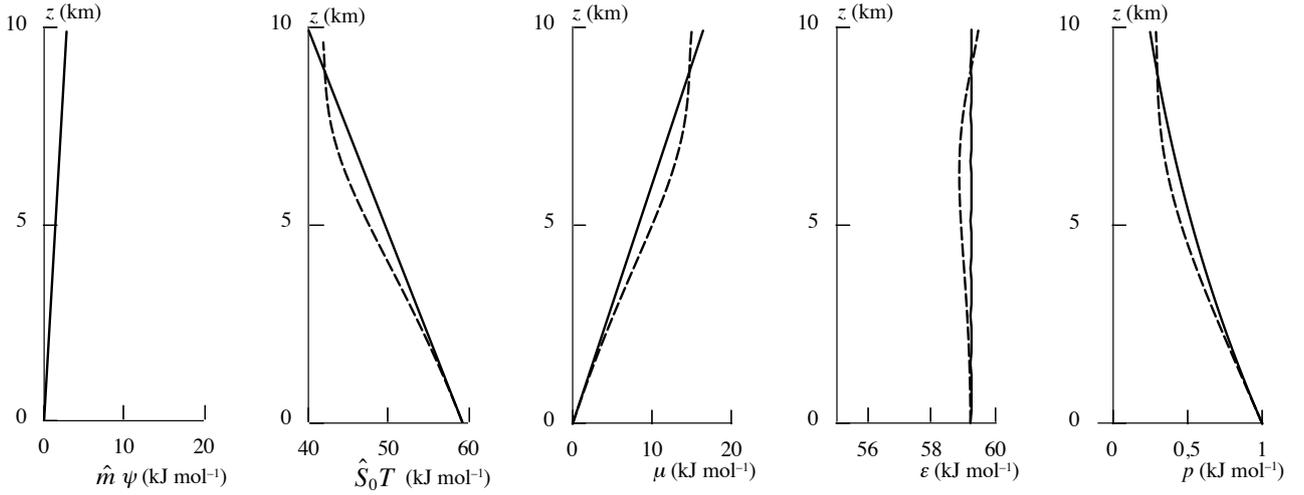

Fig. 2. The adiabatic atmosphere. The dashed lines correspond to an initial state in which the GTC equilibrium is perturbed. The full lines represent the final equilibrium. In this state of GTC equilibrium, the gravitational and the chemical driving forces balance the thermal driving force. (Scales are not the same as in Figure 1.)

Again, we have found an "old friend". If, in the case of GTC equilibrium, we divide $\varepsilon(z)$ by $c_p$ we get what in the meteorological literature is called the *potential temperature*.

Let us assume $p_0$ and $T_0$ to be the pressure and the temperature values for $z = 0$. Moreover, we shall call $\hat{S}(p_0, T_0) = \hat{S}_0$ and choose $\mu(p_0, T_0) = 0$. Using (18) equation (16) becomes

$$\mu(z) = \left( \frac{\hat{S}_0}{c_p} - 1 \right) \cdot g \cdot \hat{m} \cdot z .$$

Again, we shall write the equation with numerical values. For $\hat{S}_0$ we take 198.7 J mol⁻¹ K⁻¹ , which is the molar entropy of dry air at normal conditions, i. e., at $T = 298$ K and $p = 1$ bar:

$$\mu(z) = \frac{1.66 \text{J}}{\text{mol} \cdot \text{m}} \cdot z .$$

Just as in the case of the isothermal atmosphere, the chemical potential is a linear function of the height. Whereas for the isothermal atmosphere $\mu$ decreases with $z$, in the adiabatic atmosphere the chemical potential is an increasing function of the vertical coordinate.

We finally calculate the pressure profile of the adiabatic atmosphere by combining equation (15) with equation (18):

$$p(z) = p(0) \left( 1 - \frac{g\hat{m}}{c_p T_0} \cdot z \right)^{c_p/R} .$$

Figure 2 resumes our results graphically. The dashed lines correspond to an initial state in which the GTC equilibrium is perturbed. The full lines represent the equilibrium profiles. Notice that in the case of GTC equilibrium, the gravitational and the chemical driving forces point downwards. Both together are balanced by the thermal driving force which points upwards.

## C. Static equilibrium

In order to understand the dry atmosphere we have to discuss one more equilibrium. Let us first consider an "air column" within a closed tube, Figure 3. At the beginning, we prevent the air from flowing by means of a barrier which can be removed later. We also assume that there is no other convection within the tube. We first impose an arbitrary temperature profile over the length of the tube. We thus get a well-defined pressure profile. We then open the barrier. The air will begin to flow in the direction of the tube but, after a while, due to the friction at the walls, it will come to rest. A new equilibrium has established which is characterized by the fact that the center of mass of the air column has reached its lowest position. As this process goes on, the hydrostatic equilibrium has always acommodated itself. We call the final state the static equilibrium. Sometimes, also this equilibrium is called hydrostatic equlibrium. We remind, that the equilibrium defined by equation (12) is an equilibrium of another kind. That is why we give it another name.

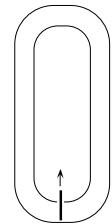

Fig. 3. If an arbitrary temperature profile (in the direction of the tube) is imposed on the air within the tube and then the barrier is opened, the air will relax into a state, in which its center of mass is at its lowest point.

Let us now study the establishment of the static equilibrium for different initial temperature profiles. We shall, however, always choose this profil identical in both vertical legs of the tube. Thus, when opening the barrier, we have static equilibrium in any case, but this equilibrium can be stable, unstable or neutral. It is neutral if

$\hat{S}(z) = \text{const},$

i. e. if the air is in GTC equilibrium. In this case, a displacement of the air column in the direction of the tube doesn't cause any change in the temperature or pressure profile.

If the temperature decrease with $z$ is steeper than according to equation (18), the static equilibrium, and thus, the "stratification", is stable. If the slope is more gentle, the stratification is unstable. If the temperature profile matches equation (18), the stratification is neutral. Of course, one can have temperature profiles which cause the stratification to be stable for certain intervals of the vertical coordinate $z$ and unstable or neutral for others. We remember, that by mixing the atmosphere, one produces conditions in which the GTC equilibrium establishes itself. We now see, that mixing always causes a neutral stratification.

### D. Equilibria of the atmosphere by day and by night

We have seen that turbulence drives the atmosphere towards a state with

$\hat{S}(z) = \text{const}.$

In this state, the temperature gradient is not equal to zero, i. e. a thermal driving force exists. If there was no turbulence, this thermal driving force would become effective. A hypothetical atmosphere, which does not exchange energy neither with the earth nor with the cosmos, would, after a long time, move into thermal equilibrium. Eventually, the atmosphere would reach the thermal and the hydrostatic equilibrium, but not the GTC equilibrium. The corresponding stratification would be stable. Actually, our real atmosphere does not reach this state, but it tries to do so every night, with some success. During the night (if the sky is clear) the surface of the earth cools more rapidly than the air. (The reason is that the surface of the earth behaves more like a black body than the air.) As a consequence, the air near to the ground cools more rapidly than the air of the higher layers of the troposphere. This causes a deviation from the GTC equilibrium corresponding to a stable stratification. The situation is very different during the day. At day-time the atmosphere is heated from below, thereby getting in a state of an unstable stratification. As a result, the atmosphere tends to turn over and induce turbulences, thereby coming back to the condition for GTC equilibrium.

## V. CONCLUSIONS

We are taught in thermodynamics that one has to distinguish various types of equilibrium. In the simplest case, to every extensive quantity belongs one type of equilibrium. If two or more extensive quantities are tightly coupled, combined equilibria will establish, and it is reasonable to define combined potentials.

We have shown, that

– the isothermal atmosphere is characterized by a constant gravito-chemical potential;

– the adiabatic,well-mixed atmosphere is characterized by a constant gravito-thermo-chemical potential.

Thus, the linear decrease of the temperature with the vertical coordinate corresponds to a state of equilibrium. The fact that the real temperatures often substantially deviate from this temperature profile is due to a permanent perturbation of this equilibrium.